# Desorption kinetics and interaction of Xe with single-wall carbon nanotube bundles

Hendrik Ulbricht[1], Jennah Kriebel[2], Gunnar Moos[1], and Tobias Hertel[1*]

[1]*Fritz-Haber-Institut der Max-Planck-Gesellschaft, Faradayweg 4-6, D-14195 Berlin, Germany*
[2]*Harvard University, Department of Chemistry and Chemical Biology, 12 Oxford St., Cambridge, MA 02138, USA*

**Abstract**

We present a study on the kinetics of xenon desorption from single-wall carbon nanotube (SWNT) bundles using thermal desorption spectroscopy (TDS). TD-spectra from SWNT samples show a broad desorption feature peaked at significantly higher temperature than the corresponding low-coverage desorption feature on graphite. The observations are explained using a coupled desorption-diffusion (CDD) model, which allows the determination of the low-coverage Xe binding energy for adsorption on SWNT bundles, 27 kJ/mol. This energy is about 25% higher than the monolayer binding energy on graphite, 21.9 kJ/mol. By comparison with molecular mechanics calculations we find that this increase of the binding energy is consistent with adsorption in highly coordinated groove-sites on the external bundle surface.

## 1. Introduction

The interaction of single-wall carbon nanotubes (SWNTs) with their environment, especially with gases or dopants adsorbed on their interior or exterior surfaces, has attracted increasing attention due to the anticipated influence on key properties of these materials [1,2]. The electronic structure and consequently the electronic transport properties of SWNTs are expected to be susceptible to the presence of adsorbates due to the fact that every atom in a SWNT can be considered a surface atom and is exposed to the outside world. This sensitivity may enable the application of SWNTs as chemical sensors. Gas adsorption on SWNTs can also find use as gas storage material, as 'containers' for chemical reactions, or as catalyst substrate. The adsorption of Xenon, which serves as model system for the interaction of inert gases with SWNT bundles, has recently been studied experimentally and theoretically by a number of groups [3,4,5,6,7].

SWNT bundles are a commonly studied form of nanotube samples [8], which consist of SWNTs arranged in a quasi-crystalline hexagonal lattice by mutual van der Waals attraction. Bundle diameters are typically a few tens of nanometers and consist of a few hundred SWNTs. Gas adsorption can occur both on the internal and external bundle surfaces. Possible binding sites include the so-called groove- and crest-sites on the external bundle surface, the endohedral sites inside of the SWNTs, and the threefold-coordinated interstitial channels located between SWNTs in the hexagonal bundle lattice. To determine which of these binding sites are preferentially occupied by an inert adsorbate, we present a detailed study on the kinetics of Xe desorption from SWNT samples using thermal desorption spectroscopy (TDS) in combination with a model of the kinetic processes involved in desorption. A newly proposed model, which we use for the analysis of thermal desorption (TD) spectra is based on coupling the desorption from the sample surface and the diffusion within the bulk of the sample. This competition between processes was unaccounted for in a previous study [3] but is essential for the interpretation of TD spectra from SWNT samples.

* Corresponding author. Fax: +49 30 8413-5383; e-mail: hertel@fhi-berlin.mpg.de

The experimental low-coverage binding energy of 27 kJ/mol is compared with molecular mechanics (MM) calculations of the energetics for different binding sites.

## 2. Experiment

The SWNT-paper samples, made from a commercially available nanotube suspension (tubes@rice, Houston, Texas) containing SWNTs with a diameter distribution peaked at 12 Å, were fabricated according to the procedure described elsewhere [8]. From the sample thickness of about 15 µm, the sample density was estimated to be 0.6 g/cm$^3$, about 60% less than the calculated density (1.5g/cm$^3$) of a close packed (9,9)- SWNT crystal. This disparity implies that approximately 60% of the sample volume is empty space, in the form of pores and voids between bundles.

Samples were outgassed by repeated heating and annealing cycles under ultra high vacuum (UHV) conditions with peak temperatures of 1200 K. This procedure ensured that traces of solvent, carboxylic groups, or other functional groups left from the purification procedure were removed prior to adsorption experiments. The SWNT-paper was attached to a tantalum disk (1 cm diameter) by adhesive forces after wetting the sample and substrate with a droplet of ethanol. A sample of highly oriented pyrolytic graphite (HOPG) was mounted with silver paint on the backside of the Ta-disk to facilitate good adhesive and thermal contact. The HOPG sample was cleaved immediately before being transferred into the vacuum chamber. The sample temperature was measured using a type K-thermocouple, spot-welded to the tantalum disk. Calibration of the thermocouple was achieved using thermal desorption spectra of thick Xe films adsorbed on the HOPG sample in combination with the heat of Xe evaporation of 15.4 kJ/mol [9,10].

The sample holder was attached to a He-continuous flow cryostat, which enabled sample cooling down to 30 K. UHV with a base pressure of $1 \cdot 10^{-10}$ mbar was maintained by a membrane, a turbo-drag and a turbo-molecular pump. Additional experimental details have been published previously [11].

Xenon of 99.99% purity was admitted to the chamber through a retractable pinhole-doser with a 10 µm diameter orifice. A constant and homogeneous flow of gas could be released onto the sample from a gas reservoir, which was kept at a pressure of typically 1 mbar. The gas flux onto the sample was on the order of $10^{-11}$ mol s$^{-1}$ cm$^{-2}$, or about 0.01 ML s$^{-1}$, where a ML (monolayer) is the concentration corresponding to a close packed Xe layer with $6.36 \cdot 10^{14}$ atoms cm$^{-2}$ or 1.06 nmol cm$^{-2}$ (1nmol=$10^{-9}$ mol) [12]. The exposure to Xe was varied over several orders of magnitude from about 0.1 L to about $10^3$ L, where 1 Langmuir (L) corresponds to an exposure to $10^{-6}$ torr s. Calibration of absolute Xe coverages was achieved using the integral of thermal desorption traces from the first saturated Xe-monolayer on HOPG samples (Fig. 1).

## 3. Results and discussion

Thermal desorption spectra from HOPG and SWNT samples that were exposed to amounts of Xe ranging from 0.1 L up to about 20 L are reproduced in Fig. 1. The sample was heated, starting at 30 K, at a rate of 0.5 Ks$^{-1}$. Thermal desorption traces from graphite (Fig. 1a) show the typical high and low temperature features attributed to desorption from the first, second and higher monolayers. The shape of these desorption features, with the exponentially increasing leading (low-temperature) edge and the sudden high-temperature cut-off, are characteristic of zero order desorption kinetics ($n$=0) according to the general rate equation

$$\frac{d\Theta}{dt} = -\nu \, \Theta^n \exp\left(-\frac{E_B}{k_B T}\right) \qquad (1)$$

where $n$ is the order of desorption, $\nu$ is the pre-exponential frequency-factor and $E_B$ is the binding energy. At low temperatures the 1$^{st}$ monolayer of an inert gas is known to grow by condensation of two-dimensional (2D) solid islands from a coexisting 2D gas [13]. The origin of zero order kinetics for desorption from graphite surfaces has been studied in detail and is attributed to the thermal equilibrium of

Xenon atoms desorbing from the 2D gas on free patches of the surface with Xe atoms desorbing from the second layer on top of the Xe islands [14,15].

The binding energy and frequency factor for desorption of the first Xe ML from HOPG, obtained from the slope of desorption traces in a log($d\Theta/dt$) vs $1/(k_B T)$ plot, are $(21.9\pm0.2)$ kJ/mol and $3.2\cdot10^{13\pm0.2}$ s$^{-1}$. This result agrees with previous studies, which obtained values between 21.8 kJ/mol and 25.2 kJ/mol [15,16].

The desorption maxima in TD traces from SWNT samples are systematically shifted to higher temperatures relative to those on HOPG, and show no saturation at similar exposures. Small shifts of the desorption peak maxima to higher or lower temperatures are observed for SWNT samples fabricated on different days. The absence of saturation at coverages exceeding 100 nmol cm$^{-2}$ is attributed to the porosity of the sample, and is a consequence of diffusion from the surface into the bulk of the material. The TD spectra also do not exhibit multilayer desorption after exposures in excess of 100 L. We find that the desorption trace maxima slowly shift to slightly lower temperatures at coverages exceeding a few tens of ML.

A comparison of TD spectra from HOPG and SWNT samples (Fig. 1) show that the SWNT desorption features (typically 30 K) are broad relative to HOPG. This cannot be accounted for by the standard rate equation (1), which gives features 2-6 K wide, irrespective of the desorption order.

We have studied the kinetics of desorption with experiments in which the acquisition of TD spectra was suspended near the temperature $T_{max}$ at which the rate of desorption reaches its maximum. The sample then cooled to the starting temperature of 30 K before the TD experiment was resumed (Fig. 2). Thermal desorption resumes near the temperature where the first run was suspended. On a single crystal surface such behavior would be attributed to inhomogeneities. We will illustrate that this need not be the case for thermal desorption from porous samples, where diffusion within the sample must be considered. However, standard models for differential thermal analysis of solid-gas reactions [17] or for the analysis of thermal desorption, from zeolites for example [18,19,20] do not account for concen-tration gradients. We will develop in the following a simple one-dimensional coupled desorption-diffusion (CDD) model which allows the calculation of the desorption rate from porous samples. The CDD model combines desorption from the parts of the surface exposed to the vacuum with transport through the open inner pores of the material.

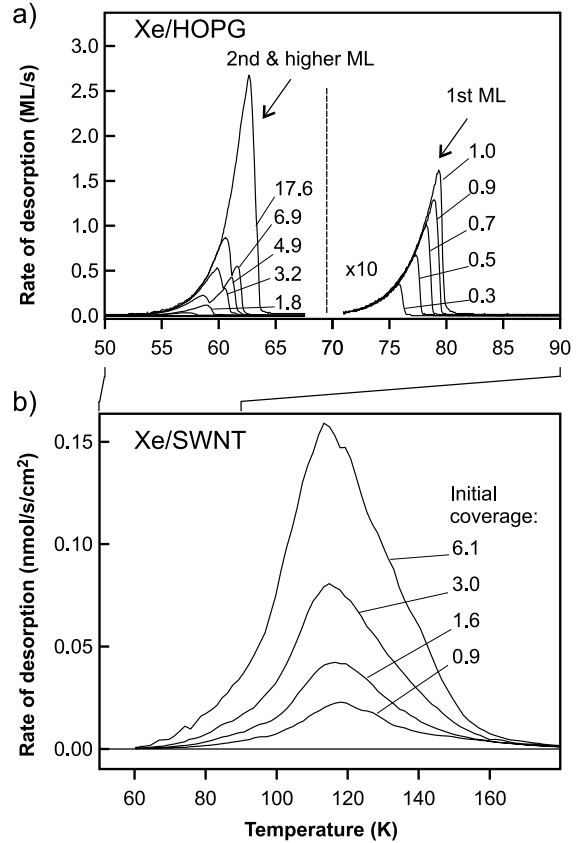

**Fig. 1**. Thermal desorption traces from graphite (upper panel) and from the SWNT sample (lower panel). The numbers next to TD traces from HOPG denote the initial coverage, which is given in units of close packed monolayers (1ML = $6.36\cdot10^{14}$ atoms cm$^{-2}$ or $1.06\cdot10^{-9}$ mol cm$^{-2}$). The initial coverage of the SWNT sample is given per geometric surface area in units of nmol·cm$^{-2}$. The heating rate was 0.5 K s$^{-1}$.

We start our discussion by introducing a $z$-dependent adsorbate concentration profile $C(z)$, where $z$ is the coordinate axis normal to the surface plane. At temperatures where Xe desorption from the bundle surfaces sets in, we expect Xe to be partially in the 'gas-phase', i.e. in open spaces and pores between the tube bundles, as well as on the bundle surfaces. These concentrations are denoted by $C_g$ and $C_s$, respectively, and are related to the total concentration $C$ by $C(z) = C_g(z) + C_s(z)$ (in mol m$^{-3}$). The evolution of the concentration profile can be calculated using a standard diffusion equation:

$$\frac{\partial C(z)}{\partial t} = \frac{\partial}{\partial z} D \frac{\partial C(z)}{\partial z} = \frac{\partial}{\partial z} j(z) \quad (2)$$

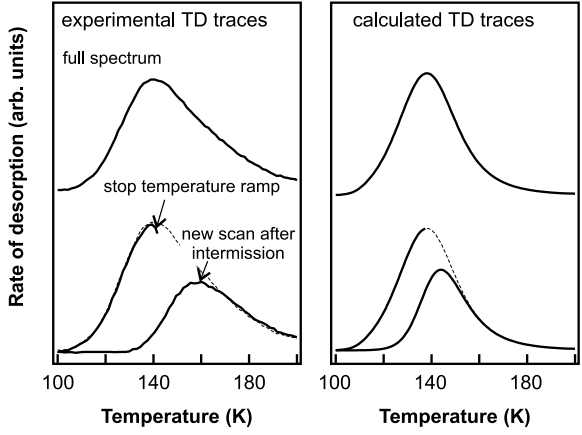

**Fig. 2**. Suspended and resumed thermal desorption traces. Experimental TD spectra (left panel) were suspended at 140 K and the sample was allowed to cool down before resuming and completing the TD experiment. The coupled desorption-diffusion model in the text allows a simple interpretation of these spectra without having to assume inhomogeneous broadening of TD features (see right panel with calculated spectra).

The total particle flux $j(z)$ is similarly separated into gas-phase and surface contributions. The flux is thus given by

$$j_x = D_x \frac{\partial C_x(z)}{\partial z}, \quad x = s, g \quad (3)$$

where $D_s$ and $D_g$ are the appropriate surface and gas phase diffusion coefficients. We rewrite the diffusion equation (2) in terms of its individual components.

$$\frac{\partial C(z)}{\partial t} = \frac{\partial}{\partial z}\left( D_s \frac{\partial C_s(z)}{\partial z} + D_g \frac{\partial C_g(z)}{\partial z} \right) \quad (4)$$

To simplify this expression, we derive a relationship between the two concentrations which assumes Langmuir kinetics [21]. This expression allows the calculation of the coverage $\Theta_t$ of the internal sample surfaces from the gas pressure $p$ using:

$$\Theta_t(p,T) = \frac{bp}{1+bp} \quad (5)$$

with

$$b = \frac{\sigma}{\nu \sqrt{2\pi m k_B T}} \exp\left(\frac{E_B}{k_B T}\right) \quad (6)$$

$\sigma$ is the surface area covered by the adsorbate and $m$ is the adsorbate mass. For $bp \ll 1$, *i.e.* for small pressure and consequently small coverages, equation (5) can be approximated by $\Theta_t(p,T) \sim bp$. This simplification is justified, since we see no significant changes in the shape of spectra for small coverages after exposure up to at least 10 L. The concentration of species on internal surfaces can be related to the surface coverage using:

$$C_s(z) = \frac{\Theta_t(z)}{N_A \sigma} A_s \quad (7)$$

where $A_s$ is the specific surface area per volume (in m$^{-1}$). $A_s$ can be estimated using the product of the surface area obtained, for example, from BET isotherms (see ref. [22]) and the SWNT-paper density. The concentration of adsorbates in the gas phase can be related to the pressure using the equation of state for an ideal gas

$$C_g(z) = \frac{p(z)}{N_A k_B T} f \quad (8)$$

rewritten as

$$C_g(z) = C(z) h(T) \quad (9)$$

with

$$h(T) = \left(1 + \frac{k_B T b A_s}{f \sigma}\right)^{-1} \quad (10)$$

$N_A$ is Avogadro's number and $f$ is the fraction of the sample volume not filled by tube bundles. These expressions can be used to simplify the gas-surface diffusion equation (4) so it reads:

$$\frac{\partial C(z)}{\partial t} = [D_g h(T) + D_s (1 - h(T))] \frac{\partial^2 C(z)}{\partial z^2} \quad (11)$$

Equation (11) is a linear diffusion equation that will be solved numerically.

We turn our attention next to the diffusion constants $D_s$ and $D_g$. The surface diffusion constant is assumed to have the usual activated form

$$D_s = D_0 \exp(-E_m/k_B T) \qquad (12)$$

$E_m$ is the rate-limiting surface migration barrier and $D_0$ is the pre-exponential factor, of the order of $10^{-7} m^2 s^{-1}$. For diffusion of gas-phase species, we adopt the simple gas-phase expression

$$D_g = \lambda v/3 \qquad (13)$$

with

$$v = (8k_B T/\pi m)^{1/2} \qquad (14)$$

where v is the mean velocity and $\lambda$ is the particle mean free path. The latter is approximated by using the mean spacing of tube bundles, equivalent to the size of the open voids. Using this expression amounts to neglecting the size of SWNT bundles and treating every gas-surface encounter in the same manner as a particle-particle scattering event in which memory of the particle momentum is lost. Here, the neglect of the finite bundle size is probably to most serious approximation.

From the simulations we find that experimental TD spectra are best reproduced if gas phase diffusion dominates the mass transport of Xe through the sample. Using eq. (11), it can be shown that this is the case if $(\lambda v/A_s D_0) \cdot \exp[(E_m - E_B)/k_B T] \gg 1$. Replacing the pre-exponential factor by $10^n$ makes this equivalent to the condition that $(E_B - E_m) < 2.3 n k_B T$. The barriers for diffusion on graphite surfaces are usually very small and are expected to be well below 1 kJ/mol. Diffusion kinetics across the outer SWNT bundle surface, however, should be strongly anisotropic with nearly vanishing barriers for migration along the bundle axis and high barriers for migration around their perimeter. Using molecular mechanics calculations we show below that the latter, rate-limiting barrier $E_m$ should be of the order of 10 kJ/mol or higher. Using the above estimatation and the fact that the pre-exponential factor $\lambda v/A_s D_0$ is of the order of $10^n$ with $n \approx 4$-5, we find that surface diffusion should dominate mass transport in these samples.

The total flux of particles leaving the sample in a TD experiment is calculated using

$$j(0) = \frac{\tilde{f}}{N_A \sigma} \frac{d\Theta(0)}{dt} + D_g \left.\frac{\partial C_g(z)}{\partial z}\right|_{z=0} \qquad (15)$$

where $\tilde{f}$ is a factor close to or slightly above one. This factor accounts for the effective increase of the visible surface area due to the roughness of SWNT samples. The first term on the righthand side of the equation is calculated from eq. (1). The second one is obtained from the concentration profile $C_g(z)$, where the evolution in time is obtained from eq. (8) after integration of eq. (11).

A calculated set of TD traces is shown in Fig. 3. For comparison, we have included traces for zero or first order desorption given by eq. (1), using the same set of parameters for each. The initial concentration was an exponentially decreasing profile with a characteristic decay length $\lambda$ of 1 µm. The evolution of this profile during the desorption experiment (see Fig. 3) can be used to explain the behavior in the suspended TD experiments described above. The behavior during the latter experiments can only be reproduced if we neglect surface diffusion altogether in agreement with the estimatation performed above. In this case the mass transport is driven by the gradient of $C_g$ and scales with the concentration $C_g$ of gas-phase species (eq. (15)). Since the gradient of $C$, and thus $C_g$, near the sample surface is reduced after ramping the temperature to $T_{max}$ (Fig. 3), desorption during the resumed temperature ramp will be negligible unless $T$ exceeds $T_{max}$. In that case the desorption continues due to the increasing concentration of gas-phase species.

These results show that the width of TD traces from porous samples as well as their shift to higher temperatures are attributed entirely to the coupled kinetics of desorption and diffusion and are not necessarily due to inhomogeneities. Occupation of a single binding site thus accounts for all of the experimental observations. Nevertheless, due to some freedom in the choice of initial conditions such as the initial concentration profile, we cannot entirely rule out that some broadening of the TD traces is caused by inhomogeneities. Evidence for an inhomogeneous contribution to binding energies can be obtained, for example, from the step width of Xe adsorption isotherms [5].

The best agreement between calculated and experimental TD spectra is obtained for a Xe binding energy of 27 kJ/mol, 25% larger than that obtained for desorption from the first monolayer on graphite. A slightly higher value of 27.2 kJ/mol was obtained in a previous isothermal adsorption study [5]. The uncertainty of the binding energy within the CDD model is estimated by computing its dependence on different parameters entering the calculation. The dependence of the binding energy $\Delta E_B/E_B$ on the pre-exponential factor v is $\Delta E_B/E_B \approx \Delta\nu\ 6\cdot 10^{-15}$ s. Using the same pre-exponential factor as for desorption of Xe from HOPG of $3\cdot 10^{13}$ s$^{-1}$ and allowing for an error in the exponent of ±0.5 we obtain an uncertainty of the binding energy of at most 3%. Further contributions may arise from the uncertainty of the mean free path within the samples $\lambda$, which should be between 20 nm and 200 nm (6% uncertainty). Another source of error could be the BET surface area, *i.e.* the area that can be covered by Xe, which would add another 3% uncertainty (here we used 300 m$^2$/g), if between 200 m$^2$/g and 500 m$^2$/g (see ref. [22]). The overall error of the binding energy resulting from a lack of information on some of the parameters within the CDD model is thus estimated to be below 15%. A more reliable determination of binding energies will require a better understanding of the kinetics of diffusion processes during gas adsorption.

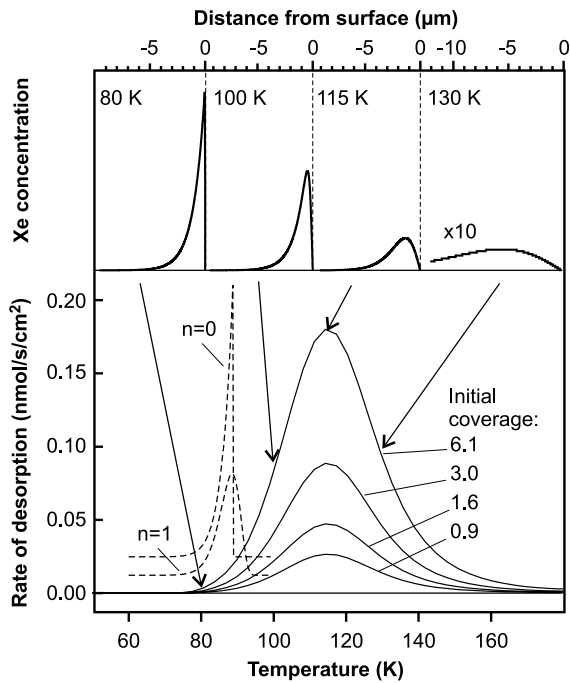

**Fig. 3**. Calculated TD traces (bottom) for different initial coverages (in ML) and concentration profiles within the SWNT sample at selected temperatures (top). The dashed lines are TD traces calculated using the same kinetic parameters for desorption from a flat surface, with n=0 and n=1, respectively.

We performed molecular mechanics (MM) simulations to determine which of the binding sites on SWNT bundles has a binding energy consistent with our experimental observations. The binding energies were computed by summation over van der Waals (VDW) pair-potentials between Xe and the underlying SWNT bundle or graphite surface using a Lenard-Jones 6-12 potential $V(r)=4\varepsilon((\sigma/r)^{12}-(\sigma/r)^{6})$. The results presented here were obtained using the VDW parameters by Stan *et al.* [7] ($\varepsilon_{Xe}$ = 221 K, $\sigma_{Xe}$ = 4.1 Å, $\varepsilon_C$ = 28 K, $\sigma_C$ = 3.4 Å).

For a comparison of calculated with experimental energies we calculate not only binding energies in the high-symmetry sites (Fig. 4a), but also mutual VDW attraction between adsorbed species. For adsorption on graphite, the calculated Xe binding energy within a 2D solid of 5.6 kJ/mol is added to the calculated binding energy of an individual Xe atom to the basal plane of 15.0 kJ/mol to obtain a total binding energy of 20.6 kJ/mol. The parameters of ref. [7] lead to an underestimation of the experimental energy, 21.9 kJ/mol, by about 6%.

The adsorption on SWNT bundles is accompanied by the formation of one-dimensional adsorbate phases, with a calculated binding energy of ~ 2 kJ/mol. The calculated binding energy for adsorption of individual Xe atoms in groove sites on

Table 1

Calculated binding energies obtained from the molecular mechanics calculations for comparison with experimental values for graphite and SWNT bundles.

| Xe binding site | Energy (kJ/mol) | |
| --- | --- | --- |
| HOPG (expt) | | 21.9±0.2 |
| Graphite | 15.0 | } 20.6 |
| 2D islands | 5.6 | |
| SWNT bundles (expt) | | 27 |
| external grooves | 21.4 | } 23.4 |
| 1D chains | 2.0 | |
| Endohedral | 24.0 | } 26.0 |
| Interstitial | < 0 | |

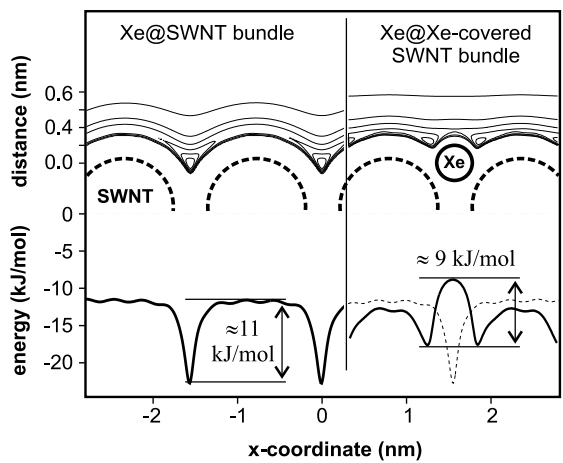

**Fig. 4**. Two-dimensional potential energy surface for a cut through a SWNT bundle near its surface. The lower part shows the minimum energy path for migration along the external bundle surface and reveals the strong lateral confinement within the groove sites as well as the high effective barriers for migration along the bundle perimeter (left: without and right: with a 1D adsorbate chain present).

the external bundle surfaces, including attractive interactions within 1D adsorbate chains, is 23.4 kJ/mol. The binding energy for adsorption in the endohedral sites is 26 kJ/mol. Taking into account the 6% underestimation of the graphite binding energy, the energies quoted above would increase slightly to ~25 kJ/mol and ~27.5 kJ/mol. The experimental value thus agrees reasonably well with both calculated binding sites. The results of these calculations are summarized in Table 1. From a determination of binding energies alone we thus cannot rule out the possibility of adsorption inside SWNTs. However, similar experiments adsorbing the larger molecules $SF_6$ and $C_{60}$ on SWNT-bundles and graphite show clearly that these gases do not adsorb in the endohedral sites, though this would be energetically the most favorable [23]. This suggests that the samples used in our experiments consist mostly of tubes where access to the internal endohedral sites is blocked. Xenon, therefore, most likely only adsorbs in the grooves on the external bundle surface.

We find that due to steric constraints the adsorption in the interstitial channels is extremely unlikely due to the energy required to expand the tube lattice to make room for the large Xe atoms.

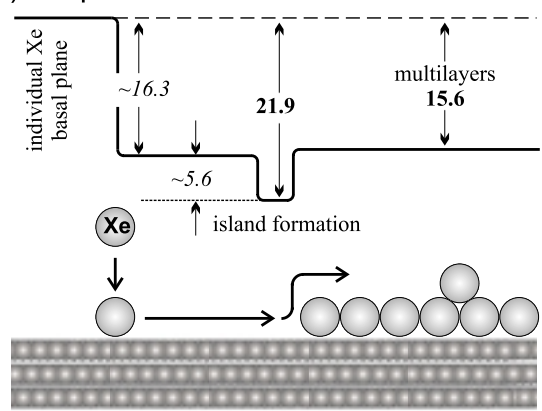

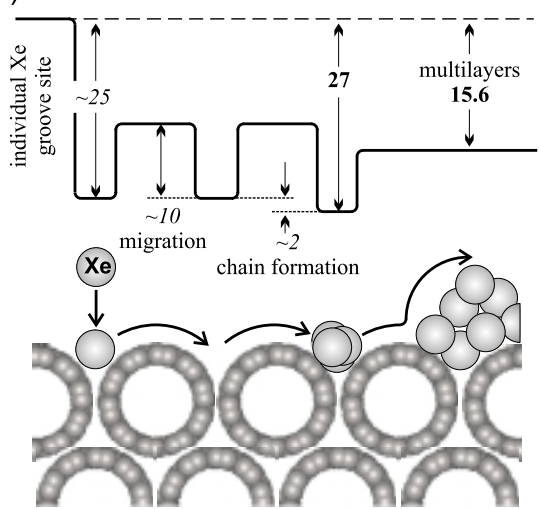

**Fig. 5**. Potential energy diagram for Xe adsorption on a) graphite and b) SWNT bundle surfaces (energies in kJ/mol). The bold numbers give experimental binding energies while italic numbers are derived from experimental energies in combination with molecular mechanics (MM) calculations or from MM-calculations alone.

A calculation of the 2-dimensional potential energy surface cut perpendicular to the bundle axis reveals the strong lateral confinement which is expected for adsorption in the external groove sites (see Fig. 4). The groove potential well is strongly confined in the lateral direction. The resulting diffusion barrier for migration around the bundle perimeter on the bare or on the Xe-covered bundle surface is nearly 10 kJ/mol. For a summary of the experimental and calculated energies obtained by this study see Fig. 5.

## 4. Conclusions

We have studied the kinetics of Xe desorption from HOPG- and SWNT-samples using thermal desorption spectroscopy. We introduced a simple coupled desorption-diffusion (CDD) model to explain and analyze the unusual shape of TD spectra from microporous SWNT samples. The model reveals that diffusion of gas phase particles through the porous structure of the sample dominates both the mass transport and desorption kinetics. The low-coverage binding energy of Xe to tube bundles is found to be 27 kJ/mol which is 25% higher than the binding energy of Xe to the basal plane of graphite of 21.9 kJ/mol. This difference can be attributed to the higher effective coordination of Xe adsorbed on the SWNT bundles and is consistent with adsorption of Xe in the grooves of the external bundle surface – as shown by a comparison with molecular mechanics calculations. For well-aligned SWNT bundles, we expect that adsorption on the external bundle surface leads to the formation of well confined nearly ideal 1-dimensional adsorbate phases.

**Acknowledgements**

We would like to thank G. Ertl for valuable comments regarding this manuscript and for his continuing support of this work.

[1] P.G. Collins, K. Bradley, M. Ishigami, A. Zettl, Science **287**, 1801 (2000)
[2] J. Kong, N.R. Franklin, C.W. Zhou, M.G. Chapline, S. Peng, K.J. Cho, H.J. Dai, Science **287**, 622 (2000)
[3] A. Kuznetsova, J. T. Yates, Jr., J. Liu and R. E. Smalley, J. Chem. Phys. **112** (2000) 9590; A. Kuznetsova, D.B. Mawhinney, V. Naumenko, J.T. Yates, Jr., J. Liu and R.E. Smalley, Chem. Phys. Lett. **321** (2000) 292.
[4] M.M. Calbi, S.M. Gatica, M.J. Bojan, and M.W. Cole, J. Chem. Phys. **115** (2001) 9975.
[5] A. J. Zambano, S. Talapatra and A. D.Migone, Phys. Rev. B **64** (2001) 075415.
[6] S. Talapatra, and A.M. Migone, Phys. Rev. Lett. **87** (2001) 206106.
[7] G. Stan, M.J. Bojan, S. Curtarolo, S.M. Gatica and M.W. Cole, Phys. Rev. B **62** (2000) 2173.
[8] A.G. Rinzler, J. Liu, H. Dai, P. Nikolaev, C.B. Huffman, F.J. Rodríguez-Macías, P.J. Boul, A.H. Lu, D. Heymann, D.T. Colbert, R.S. Lee, J.E. Fischer, A.M. Rao, P.C. Eklund, R.E. Smalley, Appl. Phys. A **67**, 29 (1998)
[9] Handbook of Chemistry and Physics, CRC ...
[10] H. Schlichting and D. Menzel, Rev. Sci. Instr. **64** (1993) 2013.
[11] T. Hertel, R. Fasel, and G. Moos, Appl. Phys. A (accepted).
[12] B. Grimm, H. Hövel, M. Pollmann, and B. Reihl, Phys. Rev. Lett. **83** (1999) 991.
[13] H. Hong, C.J. Peters, A. Mak, R.J. Birgeneau, P.M. Horn, H. Suematsu, Phys. Rev. B **40** (1989) 4797.
[14] J.A. Venables and M. Bienfait, Surf. Sci. 61 (1976) 667.
[15] J.C. Ruiz-Suarez, M.C. Vargas, F.O. Goodman and G. Scoles, Surf. Sci. **243** (1991) 219.
[16] M. Bienfait and J.A. Venables, Surf. Sci **64** (1977) 425.
[17] H.E. Kissinger, Analyt. Chem. **29** (1957) 1702.
[18] M. Sawa, M. Niwa, and Y. Murakami, Zeolites **10** (1990) 307.
[19] S.B. Sharma, B.L. Meyers, D.T. Chen, J. Miller, and J.A. Dumesic, Appl. Cat. A **102** (1993) 253.
[20] M. Niwa, N. Katada, M. Sawa, Y. Murakami, J. Phys. Chem. **99** (1995) 8812.
[21] I. Langmuir, J. Am. Chem. Soc. **40** (1918) 1361
[22] A. Fujiwara, K. Ishii, H. Suematsu, H. Kataura, Y. Maniwa, S. Suzuki, and Y. Achiba, Chem. Phys. Lett. **336** (2001) 205.
[23] J. Kriebel, H. Ulbricht, G. Moos, and T. Hertel (in preparation)